# Quantum coherence and interference of a single moiré exciton in nano-fabricated twisted semiconductor heterobilayers

*Haonan Wang[1], Heejun Kim[1], Duanfei Dong[1], Keisuke Shinokita[1],*

*Kenji Watanabe[2], Takashi Taniguchi[3], and Kazunari Matsuda[1]*

[1]Institute of Advanced Energy, Kyoto University, Uji, Kyoto 611-0011, Japan

[2]Research Center for Electronic and Optical Materials, National Institute for Materials Science, 1-1 Namiki, Tsukuba, Ibaraki 305-0044, Japan

[3]Research Center for Materials Nanoarchitectonics, National Institute for Materials Science, 1-1 Namiki, Tsukuba, Ibaraki 305-0044, Japan

**Abstract**

Moiré potential acts as periodic quantum confinement for optically generated exciton, generating spatially ordered zero-dimensional quantum system. However, broad emission spectrum arising from inhomogeneity among moiré potential hinders the exploration of the intrinsic properties of moiré exciton. In this study, we have demonstrated a new method to realize the optical observation of quantum coherence and interference of a single moiré exciton in twisted semiconducting heterobilayer beyond the diffraction limit of light. A significant single and sharp photoluminescence peak from a single moiré exciton has been demonstrated after nano-fabrication. We present the longer duration of quantum coherence of a single moiré exciton, which reaches beyond 10 ps and the accelerated decoherence process with elevating temperature and excitation power density. Moreover, the quantum interference has revealed the coupling between moiré excitons in different moiré potential minima. The observed quantum coherence and interference of moiré exciton will facilitate potential application toward quantum technologies based on moiré quantum systems.

**Introduction**

Quantum two-level system has drawn great attention in recent years due to its numerous potential applications in the field of fundamental physics, quantum simulation and quantum information processing[1–5]. Development of quantum two-level systems has led to the construction and utilization of quantum bits (qubits), which act as the fundamental unit for quantum computing and quantum information[6–8]. Resonant light-matter interactions such as Rabi oscillation[9,10], Ramsey interference[11] and Hahn echo[12] facilitate the manipulation of quantum two-level systems[13,14] by generating binary superposition states. However, the superposition states of qubits suffer from the interaction and fluctuation from the environment, leading to accelerated decoherence process[15]. The decoherence process imposes a temporal limitation on the precise manipulation of the quantum systems[16–18], causing impedance on their potential applications[19,20]. Accordingly, for platforms targeted at achieving qubits, sufficiently long coherence time is highly required[21,22]. Furthermore, it is as well necessary to control the interaction between quantum systems, since such interaction not only modifies the quantum coherence in each individual system but also enables the creation of coupled-quantum system, introducing interference or entanglement[23] between systems, which is essential for large-scale quantum devices[24–26].

The recent progress on artificial van der Waals (vdW) structures by stacking atomically thin two-dimensional (2D) materials have provide us the opportunities for the design of novel quantum platform[27–30]. The vdW heterobilayers assembled from monolayers of semiconducting transition metal dichalcogenide have exhibited the intriguing various physical phenomena, including strongly correlated insulator phases[31,32], superconductivity[33], and novel ferromagnetism[34]. Moiré superlattices with varying atomic registries in the vdW heterobilayers can be constructed by monolayer semiconducting transition metal dichalcogenides with a small lattice mismatch or twist angle. The resultant moiré superlattice leads to the formation of periodic ordered potential traps, which confine and spatially organize optically generated bound electron-hole pair (excitons) as periodic array of quantum two-level systems[35,36]. The trapped excitons in the moiré potential (moiré excitons) are expected to show the long quantum coherence due to limited degree of freedom as 0D quantum systems[37,38] and coupling interactions could be formed between the spatially-separated moiré potentials, leading to quantum interference of emitted photons[39]. These make the moiré exciton quantum systems not only a promising platform for achieving long coherence, but also an effective tool for exploring interactions within or between the quantum systems. However, experimentally, the important information on the quantum coherence and interference of moiré excitons in the vdW heterobilayers remains unexplored, because the overlapping of multiple emission peaks from moiré excitons in the inhomogeneously broadened spectra hinders these intrinsic insights within the diffraction limit of light.

In this study, we provide a new approach of quantum optics in semiconducting twisted heterobilayer beyond the diffraction limit of light. A single and sharp optical signal from a single exciton trapped in the moiré superlattice has been successfully demonstrated in $MoSe_2/WSe_2$ heterobilayer. The quantum coherence of a single moiré exciton in a potential is maintained during more than 12 ps at low temperature of 4 K, which is much longer than that of a 2D-exciton in the monolayer semiconductor [40–42]. Quantum beats with period around 230 fs are observed, indicating the coupling between moiré excitons trapped in different potentials. Furthermore, the mechanism on the quantum decoherence of a single moiré exciton state with elevating temperature and mediated by moiré exciton-exciton interaction will be discussed.

**Results and Discussion**

Fig. 1a and Fig. 1b show the schematic of the concept using nanostructure fabrication process in this study. The nano-scale fabrication using reactive ion etching (RIE) enables emissions from a single moiré exciton and observation of its quantum coherence in the $MoSe_2/WSe_2$ heterobilayer. During usual optical measurements, it is difficult to obtain clear spectra from moiré excitonic states due to the inhomogeneity of the moiré potential, resulting in ensemble-averaged and broadened emissions composed of multiple peaks, because the focused laser light with a spot size of ~1.5 μm is determined by the diffraction limit of light, which excites large number of moiré potentials due to much smaller spatial period of moiré potentials. To address this, in the microfabrication process, we apply the nano-scale fabrication technique to reduce the optical excitation and detection area of the $MoSe_2/WSe_2$ heterobilayer with a micro-pillar structure. The nanofabricated heterobilayer with a smaller pillar size than wavelength of light will provide a reduced number of spectral peaks beyond the diffraction limit of light. It is anticipated that the emission signal from a single moiré potential can be observed, which also enables to reveal the quantum coherence of moiré exciton.

The details of RIE procedures for fabrication of nanostructure is described in Methods[43]. The designed pattern for the nanostructure fabrication is shown in Fig. 1a. The black circle areas with a diameter of 2 μm, which are larger than the focused laser spot size are exposed to reactive ions and etched. The inner white areas are protected by electron-beam resist during the RIE process, and the resultant pillar sizes ($D$) corresponding to the optical active size in the heterobilayer are designed as to be 50, 100, 150, 200, and 500 nm. The typical scanning electron microscopy (SEM) image of fabricated nanostructure with an actual pillar size of 240 nm in the heterobilayer is presented in Fig. 1b. The dotted circle at the center corresponds to the pillar region in Fig. 1a, while the area between outer and inner dotted circles denotes the etched region. The SEM images of different pillar sizes are presented in Supplementary Fig. 3.

Fig. 1c shows the photoluminescence (PL) spectra in nanostructure fabricated $MoSe_2/WSe_2$ heterobilayer with various pillar sizes at 4 K. The PL spectrum in the $MoSe_2/WSe_2$ heterobilayer with pillar size ($D$) of 500 nm shows inhomogeneously broadened and ensemble averaged of multiple peaks from moiré excitons, which is consistent with the previously reported results[44]. The broadened spectrum followed by Gaussian distribution function might come from the inhomogeneous of moiré potentials in the heterobilayer. The number of peaks in the spectra are significantly decreased with decreasing the pillar size. Consequently, the PL spectrum in the heterobilayer with a pillar size of 50 nm shows a single emission peak from a moiré exciton due to a reduced small number of moiré potentials in the optical excitation and detection area determined by the pillar size. The additional series of PL spectra at the different pillars in the heterobilayer are shown in Supplementary Fig. 2a, which also shows the similar results in Fig. 1c.

Fig. 1d shows the integrated PL intensity from moiré exciton as a function of actual pillar size. With decreasing the pillar size, the integrated intensity is rapidly reduced, especially under smaller size. Owing to the Gaussian distribution of the excitation laser, power density within various pillar size is different, as shown in Supplementary Fig. 2b. The integrated intensities are calibrated with the averaged laser intensity. The result shows linear decrease with the pillar size, which also strongly supports the reduction of number of optically excited and detected moiré potentials in the microstructure fabricated $MoSe_2/WSe_2$ heterobilayer (See Supplementary Fig. 2c).

Fig. 2a shows the contour plot of excitation power dependence of normalized PL spectra from 0.8 to 3100 $W/cm^2$ in the $MoSe_2/WSe_2$ heterobilayer with $D = 50$ nm at 4 K. The power densities are calibrated based on the pillar size and the laser spot size, as described above. The spectral shape of PL spectra changes depending on the excitation power density. Fig. 2b shows the PL spectra normalized by the excitation power density. In the lower excitation power density at about 0.8 $W/cm^2$, a single and clear sharp peak from a moiré exciton with a linewidth defined by the full width at half maximum (FWHM) of 600 μeV is observed at 1.380 eV, where the spectral linewidth is almost limited by the spectral resolution of the system ($\approx$ 600 μeV). With increasing the excitation power density, the normalized PL peak of a moiré exciton at 1.380 eV gradually decreases and additional spectral peaks appears accordingly.

Fig. 2c presents the excitation power dependence of PL spectra in the expanded energy scale to clearly see the change of spectra. Only a single PL peak at 1.380 eV is observed in the lower excitation power density below 7.8 W/cm$^2$, which is benefit from nanostructure fabrication to limit the observed number of moiré potentials. With further increasing the excitation power density, the primary PL peak at 1.380 eV gradually broadened and shows the saturation behaviour around 15.6 W/cm$^2$, which will be discussed later. Moreover, the additional PL peak appears at the higher energy side of primary PL peak at 1.382 eV above 15.6 W/cm$^2$.

In order to assign the origin of spectral peaks, the PL intensity of peak as a function of excitation power density is plotted in Fig. 2e. The PL intensity at 1.380 eV linearly increases and gradually saturates around around 15.6 W/cm$^2$, which is consistent with the spectral behaviors in Fig. 2d. The saturation behavior and its saturation power density well correspond to the generated exciton density of $1.2 \times 10^{11}$ cm$^{-2}$, which is well consistent with the moiré potential density of $2.8 \times 10^{11}$ cm$^{-2}$ in the MoSe$_2$/WSe$_2$ heterobilayer with a twist angle of 1 degree. These results support the primary PL peak at 1.380 eV comes from the recombination of moiré exciton. The PL peak at 1.382 eV appears after saturation of moiré exciton emission above 15.6 W/cm$^2$ and nonlinearly increase of PL intensity is clearly obeyed the square dependence as a function of excitation power density in Fig. 2e, which suggests that the PL peak is attributed to the recombination of moiré biexciton confined in the potential. Moreover, the blue-shifted emission of moiré biexciton than that of moiré exciton due to dipolar repulsion between excitons and the energy difference of 2 meV corresponding to the binding energy of moiré biexciton are consistent with the previously reported results[45].

Fig. 2d shows the spectral linewidth of moiré exciton peak as a function of excitation power density, obtained from Voight function fitting procedure. The spectral linewidths show the narrow value as 600 μeV at lower excitation condition below 10 W/cm$^2$, where the PL intensity increases linearly. The spectral linewidth of moiré exciton peak becomes increasing depending on the excitation power density above the saturation power density of 15.6 W/cm$^2$, which suggest the influences of moiré exciton coherence depending on the exciton density.

In order to quantify the intrinsic spectral broadening, the time-dependent PL spectra in the MoSe$_2$/WSe$_2$ heterobilayer are measured. Fig. 3a shows the time evolution of PL spectra of moiré exciton, where each spectrum is accumulated during 30 seconds. The randomly changes of spectral peak positions, i.e. spectral wondering or spectral jittering are clearly observed in the moiré exciton emission, which is also used as the characteristic of 0D quantum systems[46]. Fig. 3b presents the trace of energy peak position from the data in Figure 3a. The frequencies of each peak position are shown as histogram in Fig. 3c. The histogram of energy positions during the spectral wondering ranges from 1.3755 to 1.3758 eV. The spectral wondering of PL peak is also observed in the different pillar of heterobilayer (See Supplementary Fig. 5)

In order to obtain the information on the quantum coherence, the first-order correlation function $g^{(1)}(\tau)$ of emission signals from moiré exciton is measured using a Michelson interferometer (Supplementary Fig. 6). Fig. 3d shows the counter map of interferometry of PL spectra as a function of delay time in 4 K under excitation power density of 14 W/cm². The amplitude of oscillation fringe between maximum and minimum intensity gradually decreases as delay time increases, indicating the process of decoherence, as presented by the temporal interferogram of Fig. 3e. The visibility $V(\tau)$ is calculated as following,

$$V(\tau) = \frac{I_{\max} - I_{\min}}{I_{\max} + I_{\min}} \tag{1}$$

where $I_{\max}$ and $I_{\min}$ denote the maxima and minima intensity in the interferometry. The visibility $V(\tau)$ (blue circle) as a function of the delay time is plotted in Fig. 3f. The visibility as a function of delay time corresponds to Fourier transform of the emission spectrum with the convolution result of extrinsic inhomogeneous and intrinsic inhomogeneous linewidth, in form of Lorentz and Gaussian function, respectively. As a consequence, the delay time dependent visibility in Fig. 3f can be fitted by the product of exponential and Gaussian function as follows [47,48],

$$V(\tau) = e^{-B_1 \tau} \cdot e^{-B_2^2 \tau^2} \tag{2}$$

where $B_1$ and $B_2$ are parameters defined in the exponential and Gaussian function. The fitted result using eq. (3) with $B_1$ (=0.14 ps⁻¹) and $B_2$ (=0.35 ps⁻¹) well reproduce the experimental results of visibility as a function of delay time. In the spectral domain, the homogeneous and inhomogeneous broadening of the spectrum can be described as $2\hbar B_1$, and $4\hbar\sqrt{\ln 2} B_2$, where $\hbar$ is the Plank constant divided by $2\pi$.

The exponential decay rate of $B_1$ in the visibility of interferometry is reciprocally related to the coherence time of quantum state as $T_2 = \frac{1}{B_1}$. The coherence time ($T_2$) of a quantum state can be directly described by the exciton lifetime and pure dephasing time with equation as

$$\frac{1}{T_2} = \frac{1}{2T_1} + \frac{1}{T_2^*} \tag{3}$$

where $T_1$ is the energy relaxation lifetime and $T_2^*$ is the pure dephasing time. The lifetime of interlayer moiré exciton in MoSe$_2$/WSe$_2$ heterobilayer is measured by the time-resolved PL spectroscopy using a time-correlated single-photon counting method (Supplementary Fig. 8a) from 4 to 14 K under the excitation power density of 56 W/cm². The PL decay profiles of moiré exciton show the longer decay time of several tenth ns, which is consistent with the previously reported results [44]. The decay curves are reproduced by exponential functions as $I(t) = A_1\exp(-t/\tau_1) + A_2\exp(-t/\tau_2) + A_3\exp(-t/\tau_3)$, where $A_1, A_2, A_3$ and $\tau_1, \tau_2, \tau_3$ are the coefficients of amplitude and decay times. The fitted results provide the values of $\tau_1$ = 14 ns, $\tau_2$ = 100 ns, $\tau_3$ = 700 ns at 4 K. With increasing the temperature, the PL decay profiles show faster, and the obtained decay times are summarized in Supplementary Fig. 8b. The estimated average lifetime of moiré exciton ($T_1$ = 385 ns) is much longer than the coherence time $T_2$. Thus, population relaxation process hardly contributes to the dephasing process. As a result, pure dephasing time of this position is evaluated to be 7.1 ps, corresponding to the homogeneous linewidth of 184 μeV.

Supplementary Fig. 9 shows the interferograms of moiré exciton emission at various temperatures from 4 to 12 K. The decay of interferograms becomes progressively faster with elevating temperature. Fig. 3g shows the temperature dependence of visibility, which are also well fitted using eq. (2) with changing the value of $B_1$, as shown in Fig. 3h. The evaluated coherence times of moiré exciton decrease significantly from 5.3 to 2.2 ps, corresponding to the increase of the homogeneous linewidth from 250 to 600 μeV. The broadening of intrinsic homogeneous linewidth with increasing of temperature affects that of PL linewidth, which is consistent with the results in Supplementary Fig. 4. The broadening of homogeneous linewidth with increasing temperature can be linearly modelled by $\Gamma(T) = \gamma_0 + \gamma' T$, where $\gamma_0$ is the residual homogeneous linewidth at zero temperature. Due to the relationship of $\Gamma_{homo} = 2\hbar / T_2$, coherence time $T_2$ can be described by $T_2 = 2\hbar / (\gamma_0 + \gamma' T)$. The solid line in Fig. 3h reproduces the experimental results under various temperature. According to the fitting result, the value of residual homogeneous linewidth is 83 μeV, corresponding to a coherence time of 16 ps, at zero-temperature limit and the linear coefficient of $\gamma'$ as a function of temperature is 43 μeV/$T$. The temperature dependent linearly increase of homogeneous linewidth suggests that the decoherence of moiré exciton is determined by the interaction of low energy acoustic-phonon modes in the heterobilayer. Moreover, the value of broadening coefficient $\gamma'$ dominated by the strength of exciton-acoustic phonon interaction is smaller than that of 2D exciton (60 μeV/$T$) in monolayer semiconductor[49]. This result implies that the exciton-phonon interaction is suppressed by the quantum confinement of moiré potential[50].

Supplementary Fig. 10 shows the interferograms of moiré exciton emission at 4 K

under various excitation power densities. The decay of the interferograms becomes faster with increasing the excitation power density. Fig. 3i shows the power density dependence of visibility, which also fits well using eq. (2) with changing the value of $B_1$, as shown in Fig. 3j. The evaluated coherence times of moiré exciton depending on the excitation power density significantly decrease from 7.1 to 2.0 ps, corresponding to homogeneous linewidth increasing from 180 to 660 μeV. The broadening of homogeneous linewidth with increasing excitation power density can be linearly fitted by $\Gamma(P) = \beta_0 + \beta' P$, where $\beta_0$ is the homogeneous linewidth under weak excitation power limit. Thus, coherence time can be modelled by $T_2 = 2\hbar / (\beta_0 + \beta' T)$, represented by the black solid line in Fig. 3j, indicating residual homogeneous linewidth of 160 μeV and coherence time of 8.2 ps under zero exciton density. The value of the broadening coefficient $\beta'$ induced by the exciton-exciton interaction is estimated to be 2.6 × 10$^{-3}$ meVcm$^2$/W, which is much smaller than that previously reported in monolayer WSe$_2$[51]. The smaller value indicates that the interaction between excitons is also drastically surpassed by the confinement of moiré potential [52].

We further investigated the quantum interference of moiré excitons from first-order correlation function of corresponding PL spectra in another pillar of the heterobilayer. Fig. 4a presents the results of PL spectra of moiré excitons under various excitation power density. At the lowest excitation power (0.8 W/cm$^2$), two peaks are observed at 1.327 and 1.309 eV respectively, indicated by the solid circle and the rectangle. With increasing excitation power density, each of the peaks show saturation behaviour independently and new peaks appear at the higher energy side. The intensities of these two peaks are plotted as a function of excitation power density in Fig. 4b. The intensities of each peak are fitted with function $I = P^{\alpha}$, shown by the solid lines. These two peaks present linearly increases at weak excitation power densities and different saturation power densities, indicating interlayer excitons trapped in moiré potentials of different depths ($M_1$ and $M_2$). The peak at 1.309 eV corresponds to deeper moiré potential ($M_2$) than the peak at 1.327 eV ($M_1$), which is caused by inhomogeneity across the optical excitation area.

A contour plot of the first-order correlation function of PL spectrum is presented in Fig. 4c, at an excitation power density of 3.2 W/cm$^2$. The interferogram of the peak at 1.327 eV is shown in Fig. 4d, in which the integrated peak intensity ($M_1$) is plotted as a function of delay time. The envelope of the interferogram as a function of delay time, as presented in Fig. 4e shows the coherence time of 12 ps, corresponding to a homogeneous linewidth of 110 μeV. Different from the previous results, the interferogram in Fig. 4d presents a distinct beating pattern consisting of multiple periods. The beating period $B_M$ is estimated to be 230 ± 10 fs, corresponding to an energy splitting of 18.0 ± 0.8 meV. According to the spectra in Fig. 4a, the peak positions of $M_1$ and $M_2$ show an energy difference of 18 meV, similar to the splitting energy evaluated from the beating periods in the time domain. Due to the fact that $M_1$ and $M_2$ are moiré excitons trapped in different moiré potential minima, it can be confirmed that the beating signal comes from the coupling of moiré excitons ($M_1$ and $M_2$). As indicated in Fig. 4f, the coupling between moiré excitons ($M_1$ and $M_2$) creates a four-level system, including the coherently coupled state $M_{12}$, two excited states ($M_1$ and M$_2$) trapped in moiré potentials, and the shared ground state (G). The emissions from the coupled state leads to the quantum interference (quantum beat) in the interferogram of M$_1$. The quantum couplings across adjacent electronic systems have been reported in quantum wells[53], colloidal quantum dots[54] and molecules[55,56]. However, we have demonstrated the new features of coupled quantum systems based on moiré excitons in the semiconducting heterobilayer, which can be extended to manipulate multiple quantum states in periodic moiré superlattice.

**Conclusion**

We have demonstrated a new strategy of nano-fabrication based on reactive ion etching in the study of quantum physics in semiconducting twisted MoSe$_2$/WSe$_2$ heterobilayer. The reduction of number of moiré potentials in the optically excitation and detection beyond the diffraction limit of light is realized by this nano-fabrication, which enables to observe optical signals from a single moiré exciton in the trapped potential. A significant single and sharp PL peak from a single moiré exciton in a potential has been successfully demonstrated, which also shows the characteristic spectral wondering in the 0D quantum system. We explored the quantum coherence of a single moiré exciton under various temperature and excitation power density, which suggests that the acceleration of decoherence due to moiré exciton-acoustic phonon and between moiré excitons interaction. Moreover, the longer duration of quantum coherence of a single moiré exciton has been revealed to be more than 12 ps at low temperature of 4 K, which is much longer than that of exciton in a monolayer semiconductor. Furthermore, we have observed quantum beat in the interferogram of moiré exciton, which proves the existence of coupling between moiré excitons trapped in different moiré potentials. The longer coherence of moiré exciton revealed in this study will provide the potential application toward quantum technologies based on moiré quantum systems.

**Data availability**

Data described in this paper and presented in the supplementary materials are available from the corresponding author upon reasonable request.

**Methods**

**Sample preparation and nano-fabrication process**

Monolayer (1L) MoSe$_2$, WSe$_2$, and encapsulated *h*-BN layers were prepared on SiO$_2$/Si substrates by mechanical exfoliation from the bulk crystals. The layer number and thickness of MoSe$_2$ and WSe$_2$ were determined by the optical image and PL spectra. The encapsulated MoSe$_2$/WSe$_2$ heterobilayer encapsulated by top and bottom *h*-BN was fabricated by a polydimethylsiloxane (PDMS)-based dry-transfer method. The top *h*-BN, monolayer WSe$_2$, and MoSe$_2$ were sequentially picked up by Poly (methyl methacrylate) (PMMA) coated PDMS stamp, then dropped onto bottom *h*-BN on SiO$_2$/Si substrate. The entire dry-transfer process was conducted in a N$_2$-filled glove box. The fabricated sample was thereafter immersed in acetone solution to remove residual PMMA. The electron beam lithography and selective reactive ion etching (RIE) using Ar gas were used for fabrication of designed nanostructures in MoSe$_2$/WSe$_2$ heterobilayer encapsulated *h*-BN layers. The selective RIE using Ar gas was conducted in the conditions of power (50 W) and flow rate (40 s.c.c.m), respectively.

**Optical measurement**

A linearly polarized semiconductor laser (2.38 eV) was used as an excitation power source for low temperature PL measurements. A 50× objective lens with N.A.=0.67 for focusing the excitation laser light on the surface and acquiring optical image was used. The sample was placed in a cryogen-free cryostat with temperature ranging from 4 K to room temperature. The emission signals were coupled with an optical fiber and detected by a charge coupled device through a spectrometer with a typical spectral resolution of 0.6 meV. A monochromatic pulsed supercontinuum light source by the bandpass filter with a photon energy of 1.7 eV and repetition rate of 1 MHz was used for the time-resolved PL measurement. The emission signals were delivered through an optical fiber and detected by a Si avalanche photodiode by a time-correlated single photon counting technique. A Michelson interferometer was used to record the first-order correlation function $g^1(\tau)$ of PL signals for coherence measurement. The detail optical setup is shown in Supplementary Fig. 6.

**Reference**

1.  Zrenner, A. et al. Coherent properties of a two-level system based on a quantum-dot photodiode. *Nature* **418**, 612–614 (2002).

2.  Veldhorst, M. et al. A two-qubit logic gate in silicon. *Nature* 526, 410–414 (2015).

3.  Monz, T. et al. 14-qubit entanglement: Creation and coherence. *Phys. Rev. Lett.* **106**, 130506 (2011).

4.  Zhe Xian Koong et al. Coherence in cooperative photon emission from indistinguishable quantum emitters. *Sci. Adv.* **8**, eabm8171 (2022).

5.  Kranz, L. et al. The Use of Exchange Coupled Atom Qubits as Atomic-Scale Magnetic Field Sensors. *Adv. Mater.* **35**, 2201625 (2023)

6.  Petersson, K. D., Petta, J. R., Lu, H. & Gossard, A. C. Quantum coherence in a one-electron semiconductor charge qubit. *Phys. Rev. Lett.* **105**, 246804 (2010).

7.  Zhai, L. et al. Quantum interference of identical photons from remote GaAs quantum dots. *Nat. Nanotechnol.* **17**, 829–833 (2022).

8.  He, Y. M. et al. Coherently driving a single quantum two-level system with dichromatic laser pulses. *Nat. Phys.* **15**, 941–946 (2019).

9.  Rabi, I. I. Space Quantization in a Gyrating Magnetic Field. *Phys. Rev.* **51**, 652-654 (1937).

10. Kamada, H., Gotoh, H., Temmyo, J., Takagahara, T. & Ando, H. Exciton rabi oscillation in a single quantum dot. *Phys. Rev. Lett.* **87**, 246401 (2001).

11. Ramsey, N. F. A molecular beam resonance method with separated oscillating fields. *Phys. Rev.* **78**,695-699 (1950).

12. Hahn, E. L. Spin echoes. *Phys. Rev.* **80**, 580-594 (1950).

13. Hildner, R., Brinks, D. & Van Hulst, N. F. Femtosecond coherence and quantum control of single molecules at room temperature. *Nat. Phys.* **7**, 172–177 (2011).

14. Cai, R. et al. Zero-field quantum beats and spin decoherence mechanisms in $CsPbBr_3$ perovskite nanocrystals. *Nat. Commun.* **14**, 2472 (2023).

15. Huang, P. et al. Observation of an anomalous decoherence effect in a quantum bath at room temperature. *Nat. Commun.* **2**, 570 (2011).

16. Yoneda, J. et al. A quantum-dot spin qubit with coherence limited by charge noise and fidelity higher than 99.9%. *Nat. Nanotechnol.* **13**, 102–106 (2018).

17. Laird, E. A., Pei, F. & Kouwenhoven, L. P. A valley-spin qubit in a carbon nanotube. *Nat. Nanotechnol.* **8**, 565–568 (2013).

18. Place, A. P. M. et al. New material platform for superconducting transmon qubits with coherence times exceeding 0.3 milliseconds. *Nat. Commun.* **12**, 1779 (2021).

19. Meier, F., Levy, J. & Loss, D. Quantum Computing with Spin Cluster Qubits. *Phys. Rev. Lett.* **90**, 047901 (2003).

20. Kamyar Saeedi et al. Room-Temperature Quantum Bit Storage Exceeding 39 Minutes Using Ionized Donors in Silicon-28. *Science* **342**, 830-833 (2013).

21. Wang, P. et al. Single ion qubit with estimated coherence time exceeding one hour. *Nat. Commun.* **12**, 233 (2021).

22. Bader, K. et al. Room temperature quantum coherence in a potential molecular qubit. *Nat. Commun.* **5**, 5304 (2014).

23. Bouwmeester, D. et al. Experimental quantum teleportation. *Nature* **390**, 575-579

(1997).

24. Mandal, S. et al. The diamond superconducting quantum interference device. *ACS Nano* **5**, 7144–7148 (2011).

25. Ke, S. H., Weitao, Y. & Baranger, H. U. Quantum-interference-controlled molecular electronics. *Nano Lett.* **8**, 3257–3261 (2008).

26. Girit, Ç. et al. Tunable graphene dc superconducting quantum interference device. *Nano Lett.* **9**, 198–199 (2009).

27. Shimazaki, Y. et al. Strongly correlated electrons and hybrid excitons in a moiré heterostructure. *Nature* **580**, 472–477 (2020).

28. Tang, K. et al. Moiré-Pattern-Tuned Electronic Structures of van der Waals Heterostructures. *Adv. Funct. Mater.* **30**, 2002672 (2020).

29. H. Baek et al. Highly energy-tunable quantum light from moiré-trapped excitons. *Sci. Adv.* **6**, eaba8526 (2020).

30. Hongyi Yu et al. Moiré excitons: From programmable quantum emitter arrays to spin-orbit–coupled artificial lattices. *Sci. Adv.* **3**, e1701696 (2017).

31. Cao, Y. et al. Correlated insulator behaviour at half-filling in magic-angle graphene superlattices. *Nature* **556**, 80–84 (2018).

32. Miao, S. et al. Strong interaction between interlayer excitons and correlated electrons in $WSe_2/WS_2$ moiré superlattice. *Nat. Commun.* **12**, 3608 (2021).

33. Cao, Y. et al. Unconventional superconductivity in magic-angle graphene superlattices. *Nature* **556**, 43–50 (2018).

34. Wang, X. et al. Light-induced ferromagnetism in moiré superlattices. *Nature* **604**, 468–

473 (2022).

35. Haider, G. et al. Superradiant Emission from Coherent Excitons in van Der Waals Heterostructures. *Adv. Funct. Mater.* **31**, 2102196 (2021).

36. Shinokita, K. et al. Valley Relaxation of the Moiré Excitons in a WSe$_2$/MoSe$_2$ Heterobilayer. *ACS Nano*, **16**, **10**, 16862–16868 (2022).

37. Moody, G. et al. Electronic Enhancement of the Exciton Coherence Time in Charged Quantum Dots. *Phys. Rev. Lett.* **116**, 037402 (2016).

38. Hendrik Utzat et al. Coherent single-photon emission from colloidal lead halide perovskite quantum dots. *Science* **363**, 1068-1072 (2019).

39. Htoon, H. et al. Interplay of rabi oscillations and quantum interference in semiconductor quantum dots. *Phys. Rev. Lett.* **88**, 87401 (2002).

40. Jakubczyk, T. et al. Radiatively Limited Dephasing and Exciton Dynamics in MoSe$_2$ Monolayers Revealed with Four-Wave Mixing Microscopy. *Nano Lett.* **16**, 5333–5339 (2016).

41. Selig, M., Berghäuser, G., Raja, A. et al. Excitonic linewidth and coherence lifetime in monolayer transition metal dichalcogenides. *Nat. Commun.* **7**, 13279 (2016).

42. Shepard, G. D. et al. Trion-Species-Resolved Quantum Beats in MoSe$_2$. *ACS Nano* **11**, 11550–11558 (2017).

43. Liu, Z. et al. In-plane heterostructures of graphene and hexagonal boron nitride with controlled domain sizes. *Nat. Nanotechnol.* **8**, 119–124 (2013).

44. Förg, M., Baimuratov, A.S., Kruchinin, S.Y. et al. Moiré excitons in MoSe$_2$-WSe$_2$ heterobilayers and heterotrilayers. *Nat. Commun.* **12**, 1656 (2021).

45. Li, W., Lu, X., Dubey, S., Devenica, L. & Srivastava, A. Dipolar interactions between localized interlayer excitons in van der Waals heterostructures. *Nat. Mater.* **19**, 624–629 (2020).

46. Kamada, H. & Kutsuwa, T. Broadening of single quantum dot exciton luminescence spectra due to interaction with randomly fluctuating environmental charges. *Phys. Rev. B.* **78**, 155324 (2008).

47. Kuroda, T., Sakuma, Y., Sakoda, K., Takemoto, K. & Usuki, T. Single-photon interferometry in InAsInP quantum dots emitting at 1300 nm wavelength. *Appl. Phys. Lett.* **91**, 223113 (2007).

48. Zhang, Y. et al. Magnon-Coupled Intralayer Moiré Trion in Monolayer Semiconductor–Antiferromagnet Heterostructures. *Adv. Mater.* **34**, 2200301 (2022).

49. Rivera, P. et al. Observation of long-lived interlayer excitons in monolayer $MoSe_2$-$WSe_2$ heterostructures. *Nat. Commun.* **6**, 6242 (2015).

50. Dey, P. et al. Optical Coherence in Atomic-Monolayer Transition-Metal Dichalcogenides Limited by Electron-Phonon Interactions. *Phys. Rev. Lett.* **116**, 127402 (2016).

51. Moody, G. et al. Intrinsic homogeneous linewidth and broadening mechanisms of excitons in monolayer transition metal dichalcogenides. *Nat. Commun.* **6**, 8315 (2015).

52. Shahnazaryan, V., Iorsh, I., Shelykh, I. A. & Kyriienko, O. Exciton-exciton interaction in transition-metal dichalcogenide monolayers. *Phys. Rev. B.* **96**, 115409 (2017).

53. Faist, J., Capasso, F., Sirtori, C., West, K. W. & Pfeiffer, L. N. Controlling the sign of quantum interference by tunnelling from quantum wells. *Nature* **390**, 589–591 (1997).

54. Cui, J. et al. Colloidal quantum dot molecules manifesting quantum coupling at room temperature. *Nat. Commun.* **10**, 5401 (2019).

55. Gerlich, S. et al. Quantum interference of large organic molecules. *Nat. Commun.* **2**, 263 (2011).

56. Gunasekaran, S., Greenwald, J. E. & Venkataraman, L. Visualizing Quantum Interference in Molecular Junctions. *Nano Lett.* **20**, 2843–2848 (2020).

**Acknowledgements**

This work was supported by JSPS KAKENHI (Grant Numbers JP16H00910, JP16H06331, JP17H06786, JP19K14633, JP19K22142, JP20H05664, JP21H05232, JP21H05235, JP21H01012, JP21H05233, and JP22K18986), JST FOREST program (Grant Number JPMJFR213K), the Keihanshin Consortium for Fostering the Next Generation of Global Leaders in Research (K-CONNEX) established by the Human Resource Development Program for Science and Technology, MEXT, and the Collaboration Program of the Laboratory for Complex Energy Processes, Institute of Advanced Energy, Kyoto University. Growth of $h$-BN was supported from Grant No. JPMXP0112101001, JSPS KAKENHI Grant No. JP20H00354.

**Author Contribution**

H.W. K.W. and T.T. contribute to the fabrication of samples studied here. H.W., K.S. and K.M. designed the experiments that were performed by H.W., H.K. and K.M. Data analysis was performed by H.W., and K.M. The draft was written by H.W.,K.S., and K.M., with all authors contributing to reviewing and editing. The project was supervised by K.M.

**Competing interests**

Authors declare that they have no competing interests.

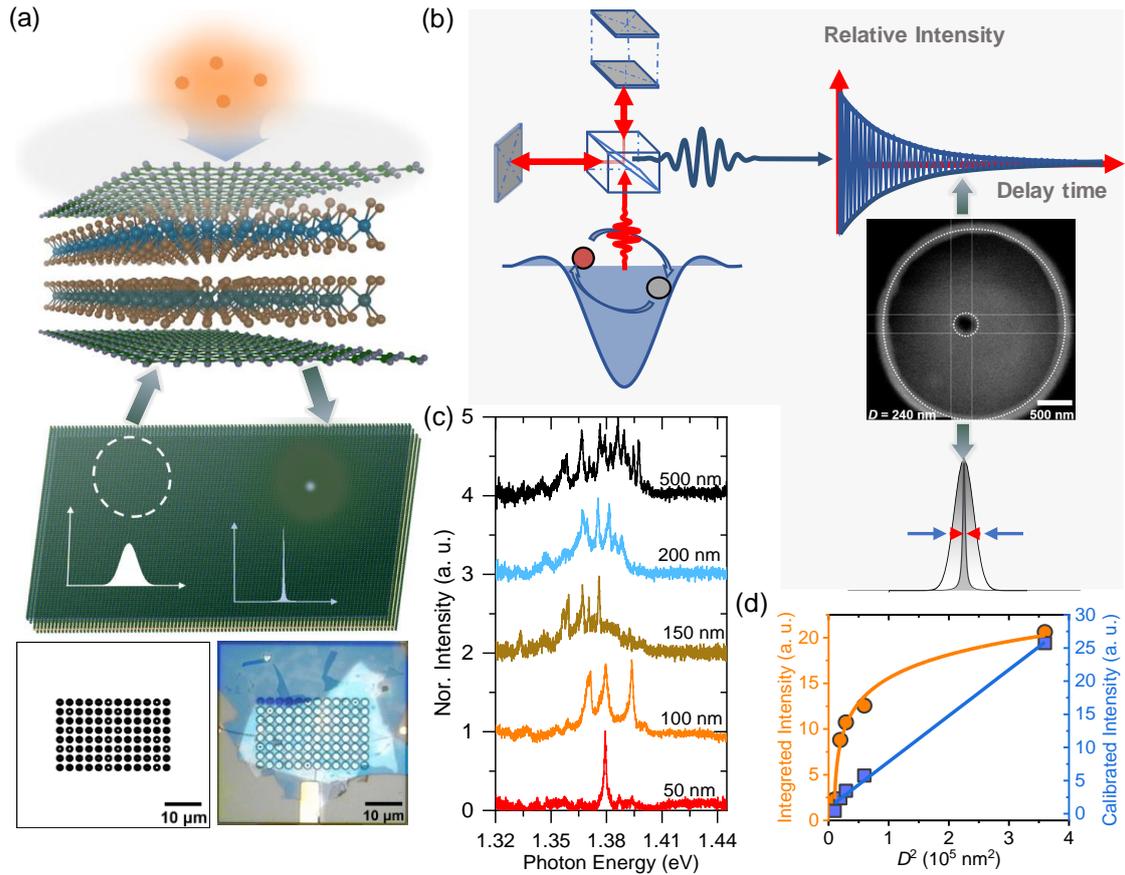

**Figure 1 (a) (b)** Schematic of the concept in this study. Nano-scale fabrication using reactive ion etching enables emissions from a single moiré exciton and observation of its quantum coherence with Michelson interferometer in the MoSe$_2$/WSe$_2$ heterobilayer beyond the diffraction limit of light. Designed pattern for RIE (left) and the optical image of MoSe$_2$/WSe$_2$ heterobilayer with an array of nanofabricated structures (right) are presented in (a). The designed sizes of pillar in the fabrications are 50, 100, 150, 200, and 500 nm. A SEM image of typical pillar is presented in (b). The inner dotted circle shows a pillar with a diameter of 240 nm corresponding to the optically excitation and observation area of moiré potential. Optical spectra form the pillars are Fourier-transformed into temporal interferograms by the Michelson interferometer. **(c)** PL spectra of MoSe$_2$/WSe$_2$ heterobilayers with various pillar sizes at 4 K. **(d)** Integrated PL intensity and calibrated intensity of heterobilayer as a function of pillar size. The solid line in the image shows the guide line, where the calibrated intensity is defined as Integrated intensity/Average laser power density. The calibrated intensity shows linearly dependence on $D^2$, indicating linearly dependence of peak numbers on the pillar size.

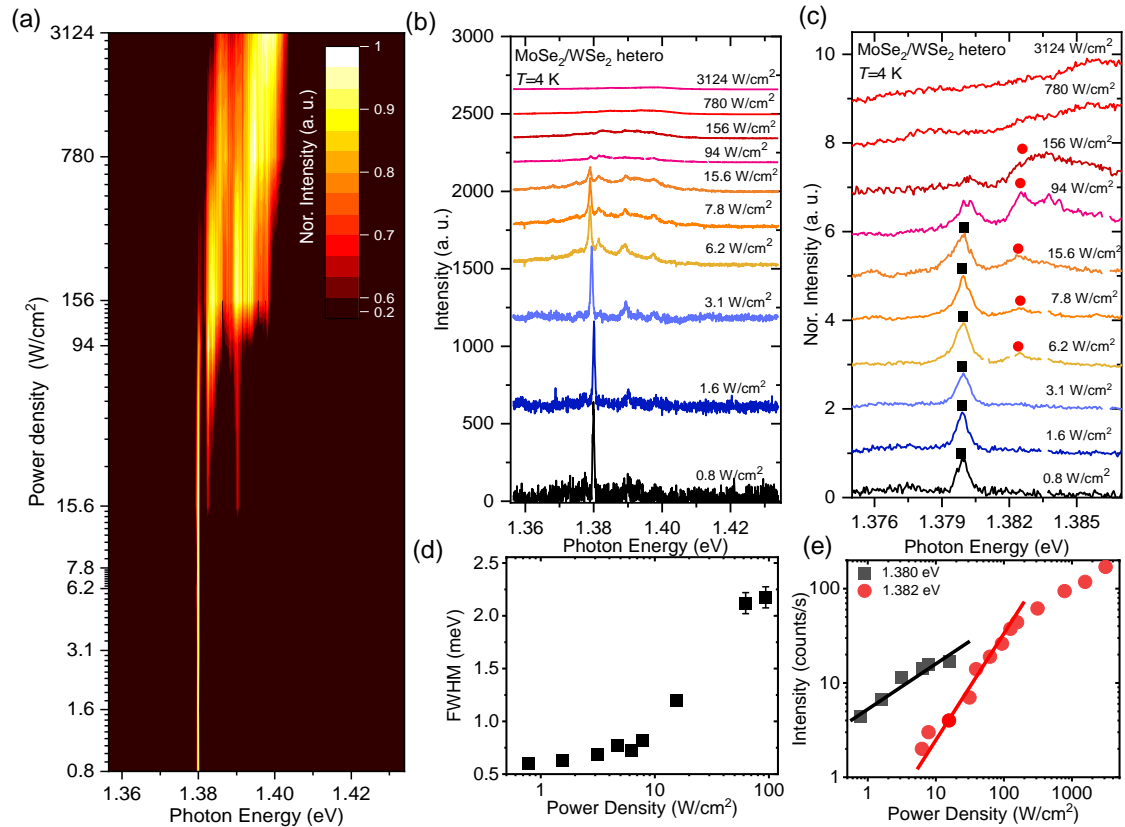

**Figure 2 Power dependence of PL spectra (a)** Contour plot of normalized PL spectra of MoSe$_2$/WSe$_2$ heterobilayer at 4 K under various excitation power densities. **(b)** Low temperature PL spectra of heterobilayer with a pillar diameter of 200 nm under various excitation power densities. **(c)** PL spectra of heterobilayer in the expanded energy scale. **(d)** Spectral linewidth of PL peak at 1.380 eV defined by full width at half maximum (FWHM) as a function of excitation power densities. **(e)** PL intensities of peaks at 1.380 and 1.382 eV as a function of excitation power density. The red and black line show the linear and square excitation power dependence.

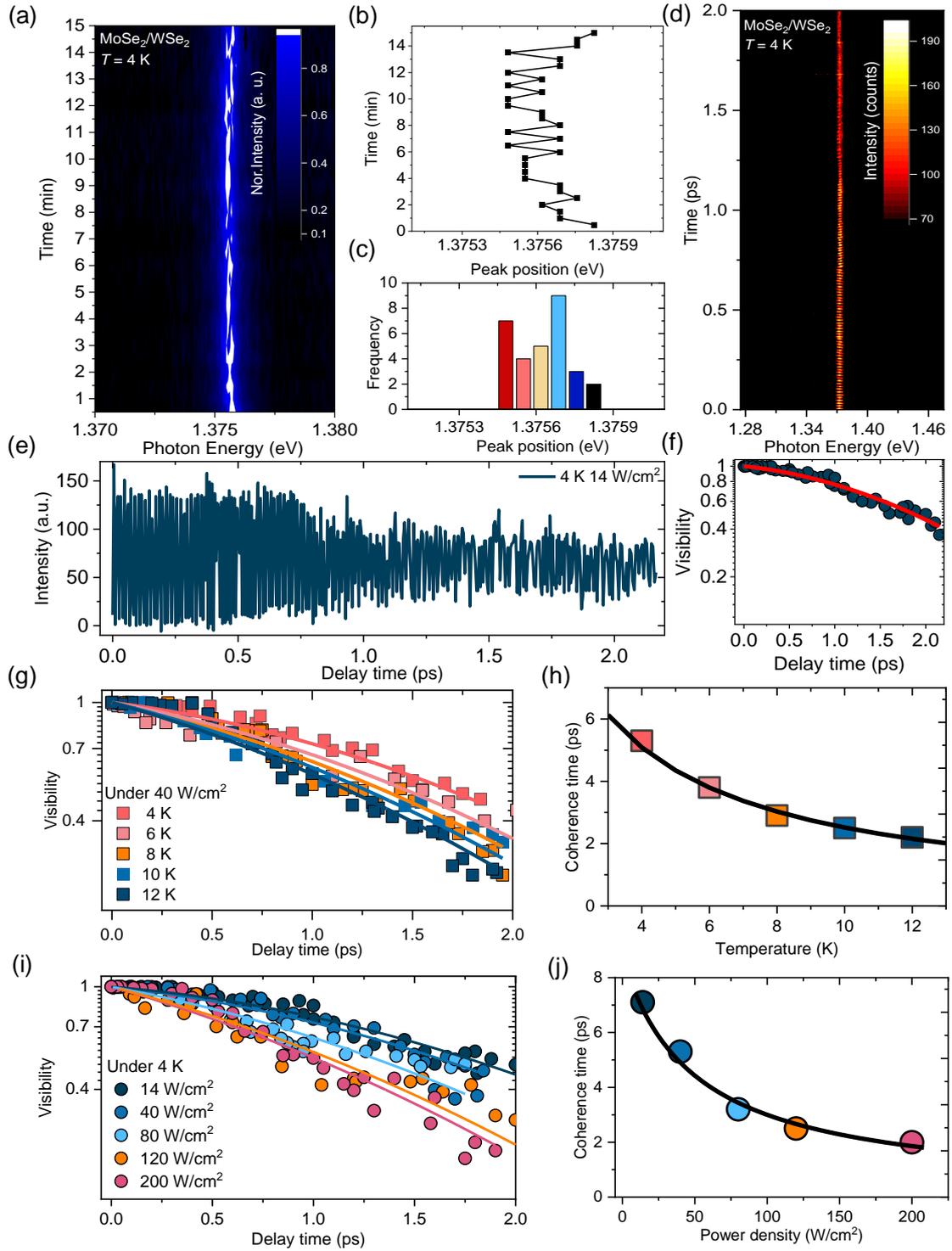

**Figure 3 PL spectral wondering and first-order correlation function (a)** Time-evolution of PL spectrum of moiré exciton peak at low temperature. Accumulation time for each spectrum is 30 seconds. **(b)** Time-trace of energy peak position of each spectrum from the contour plot. **(c)** Frequencies of energy peak positions as a histogram. **(d)** Counter plot of first-order correlation function of PL spectrum using Michelson

interferometer as a function of delay time. **(e)** Interferogram of the moiré exciton peak in the time-domain at 4 K under excitation power condition of 14 W/cm$^2$. **(f)** Decay profile of visibility in the interferogram. The solid curve shows the fitted result of the product of an exponential and a Gaussian function. **(g)** Temperature dependence of visibility of the moiré exciton peak as a function of delay time. The solid curves are fitting results of the product of an exponential and a Gaussian function. **(h)** Plot of extracted coherence time of moiré exciton ($T_2$) as a function of temperature. **(i)** Visibility of the moiré exciton peak as a function of delay time under different excitation power densities. **(j)** Plot of coherence time as a function of power density.

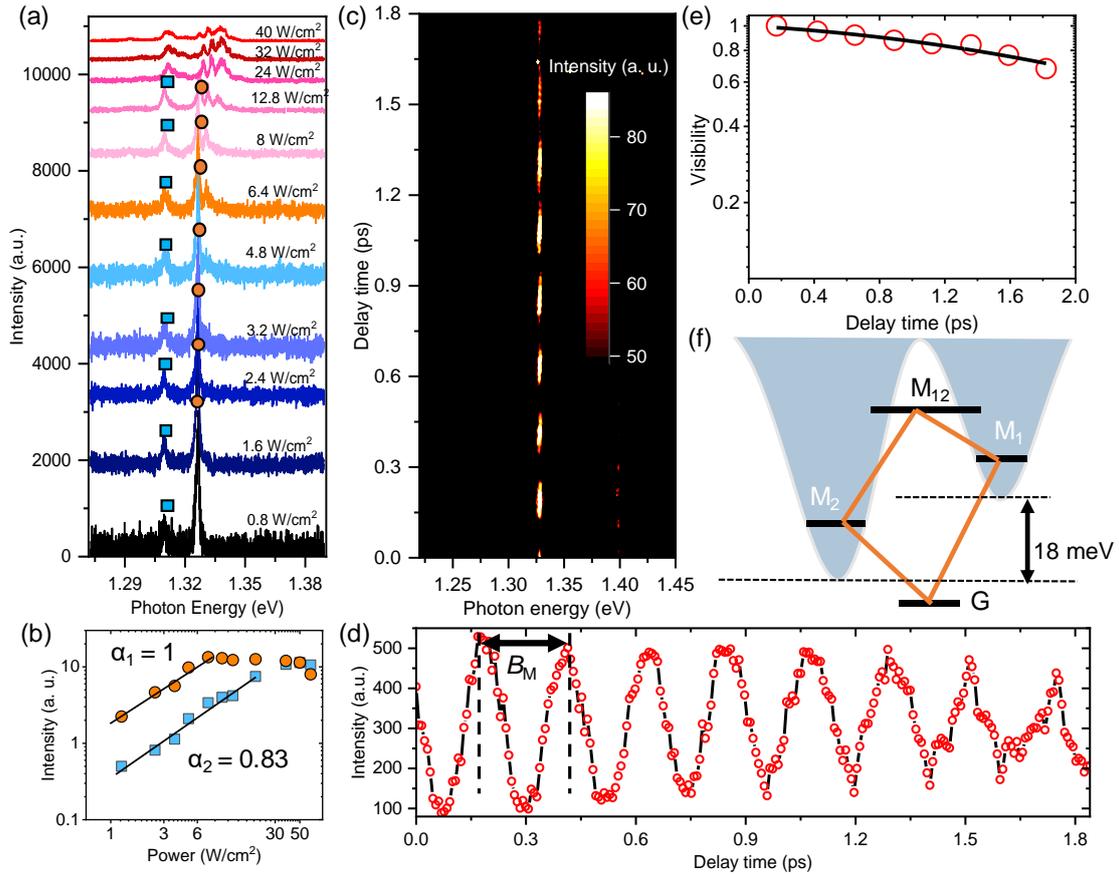

**Figure 4 Quantum beat and coupling between two moiré excitons (a)** Low temperature (4 K) PL spectra of a new pillar with diameter of 50 nm under various excitation powers. **(b)** PL intensities of peaks at 1.327 and 1.309 eV as a function of excitation power density. **(c)** Contour plot of first-order correlation function of PL spectra using Michelson interferometer as a function of delay time at the same position under excitation power of 3.2 W/cm$^2$. **(d)** Interferogram of the peak at 1.327 eV. The interferogram shows beating signal with beating period $B_M = 230 \pm 10$ fs. **(e)** Envelope of the interferogram and its fitting result. **(f)** Schematic of the coupling from moiré excitons from two different moiré potential minima. $M_1$ indicates the moiré exciton state at 1.327 eV, while $M_2$ indicates the moiré exciton state at 1.309 eV. Period of beating signal reveals the energy splitting between two moiré excitons.

# Supplementary Information

# Quantum coherence and interference of a single moiré exciton in nano-fabricated twisted semiconductor heterobilayers

*Haonan Wang[1], Heejun Kim[1], Duanfei Dong[1], Keisuke Shinokita[1],*

*Kenji Watanabe[2], Takashi Taniguchi[3], and Kazunari Matsuda[1]*

[1]Institute of Advanced Energy, Kyoto University, Uji, Kyoto 611-0011, Japan

[2]Research Center for Electronic and Optical Materials, National Institute for Materials Science, 1-1 Namiki, Tsukuba, Ibaraki 305-0044, Japan

[3]Research Center for Materials Nanoarchitectonics, National Institute for Materials Science, 1-1 Namiki, Tsukuba, Ibaraki 305-0044, Japan

**Supplementary Note 1. Estimation of twist angle from optical image**

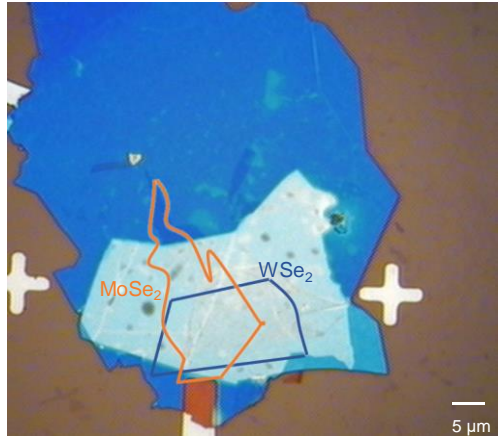

**Supplementary Figure 1 Optical image of twisted MoSe₂/WSe₂ heterobilayers encapsulated by *h*-BNs.** Solid blue and orange line correspond to monolayer $MoSe_2$ and $WSe_2$ region, respectively. The twist angle between monolayer $MoSe_2$ and $WSe_2$ is evaluated as to be about 1 degree.

**Supplementary Note 2. Estimation of the optically generated exciton density**

The exciton density is estimated under the condition of continuous-wave laser excitation. The equation for estimation is as follow:

$$\frac{dN}{dt} = g - \frac{N}{\tau_{nrad}} - \frac{N}{\tau_{rad}}, \tag{S1}$$

$$g = \frac{(1-R)\alpha dP}{\hbar\omega}, \tag{S2}$$

where $g$ represents the generation rate of moiré exciton, and $N$ is the density of moiré exciton. $\tau_{rad}$ and $\tau_{nrad}$ are the radiative lifetime and nonradiative lifetime respectively. $R$ (=0.4) is the reflectivity, $\alpha$ (=2 × 10⁵ cm⁻¹) denotes the absorption coefficient, $d$ (= 1.4 nm) is the thickness of the heterobilayer. $P$ is the excitation power density. $\hbar\omega$ (= 2.38 eV) is the excitation photon energy determined by the laser.

Under steady-state condition, decay rate equals to the generation rate, leading to:

$$N = g / (\frac{1}{\tau_{nrad}} + \frac{1}{\tau_{rad}}), \tag{S3}$$

$(\frac{1}{\tau_{nrad}} + \frac{1}{\tau_{rad}})^{-1}$ denotes the decay time obtained from experiment.

Thus, we have evaluated the exciton density of approximately $1.2 \times 10^{11}$ cm⁻², which is comparable to the density of moiré potential at twist angle of 1 degree.

**Supplementary Note 3. Calculation of laser power density**

The spatial beam profile of continuous-wave laser follows the Gaussian distribution function in the transverse direction to $z$-axis. The intensity distribution can be given by,

$$I(r, z) = I_z \exp\left(\frac{-2r^2}{W(z)^2}\right), \tag{S4}$$

where $r\ (=\sqrt{x^2+y^2})$ is the radial distance from the $z$-axis, $W(z)$ denotes the radius of the laser profile, and $I_z$ is the amplitude at $z = 0$ in the focused plane. Accordingly, the average excitation power density at given pillar size $D$ can be calculated as,

$$P_{\text{avr}} = \frac{\oint_0^D I(r)dr}{S(r)}, \tag{S5}$$

where $S(r)$ is the areal size with radius $r$, denoted as $\pi r^2$.

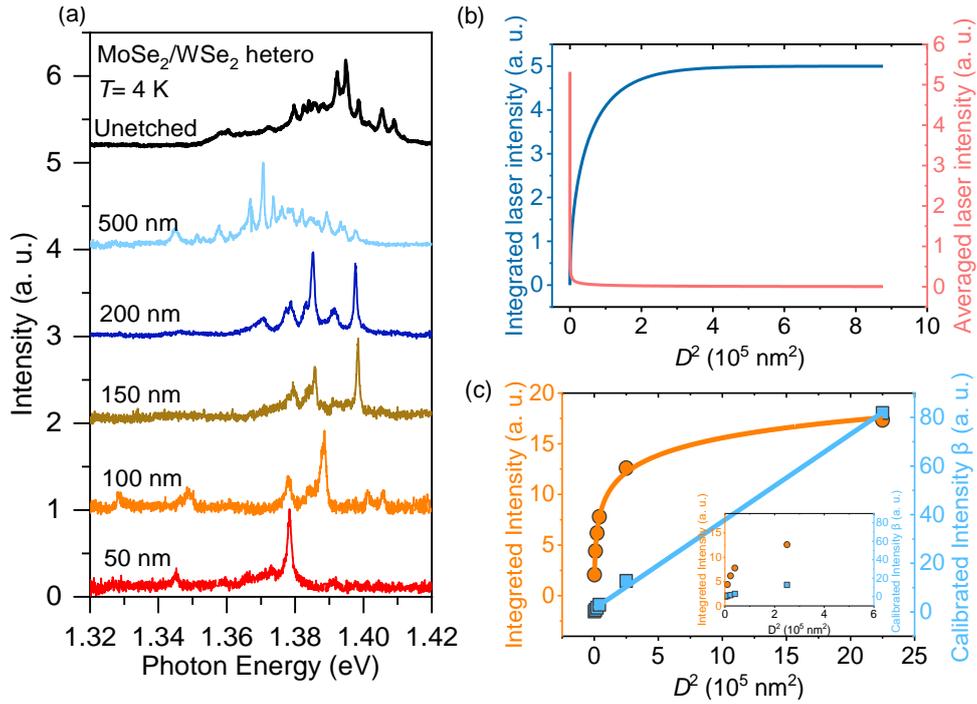

**Supplementary Figure 2: PL spectra in the nanofabricated MoSe$_2$/WSe$_2$ heterobilayer with various pillar sizes, measured at the different positions under 4 K and weak excitation conditions of 2 W/cm$^2$ (a)** Normalized PL spectra of moiré excitons in the MoSe$_2$/WSe$_2$ heterobilayer without and with nanofabrication with various pillar sizes. **(b)** Calculated integrated and averaged laser intensity as a function of square of pillar sizes, where the diameter of Gaussian profile of focused laser spot of 1.5 μm is assumed. **(c)** Integrated PL intensity as a function of squared diameter, denoted as solid circles. The blue squares show the calibrated results.

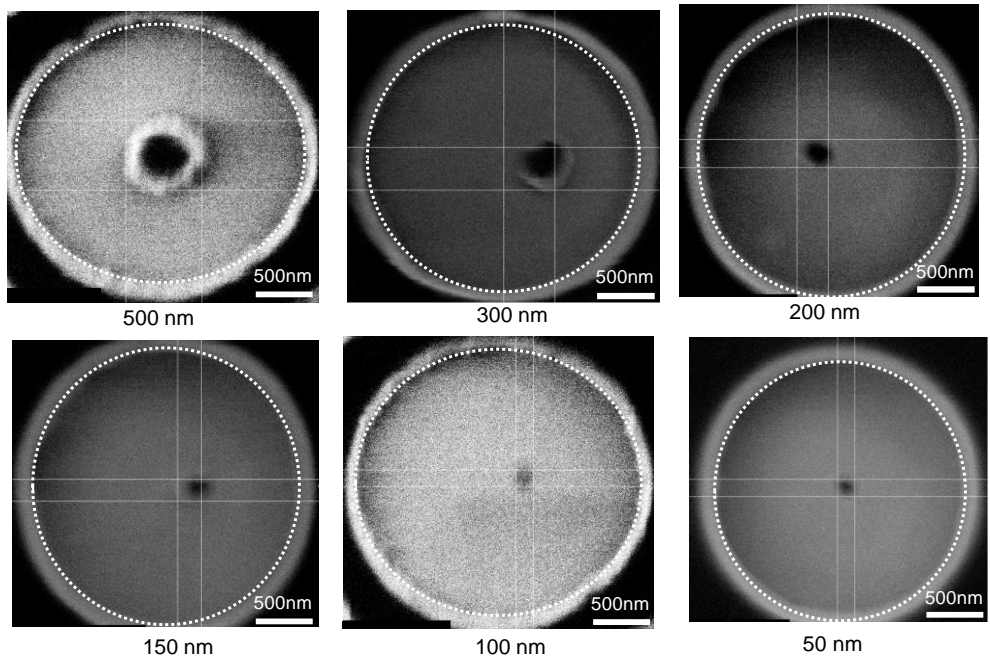

**Supplementary Figure 3: SEM images of the nanofabricated MoSe$_2$/WSe$_2$ heterobilayer with various pillar sizes.** The etched regions of outer circles are designed to be a diameter of 2 μm, indicated by the white circles in the figures. Shape and sizes of the inner pillars are determined by designed pattern by electron-beam lithography. The designed diameters of pillars are shown in the figures.

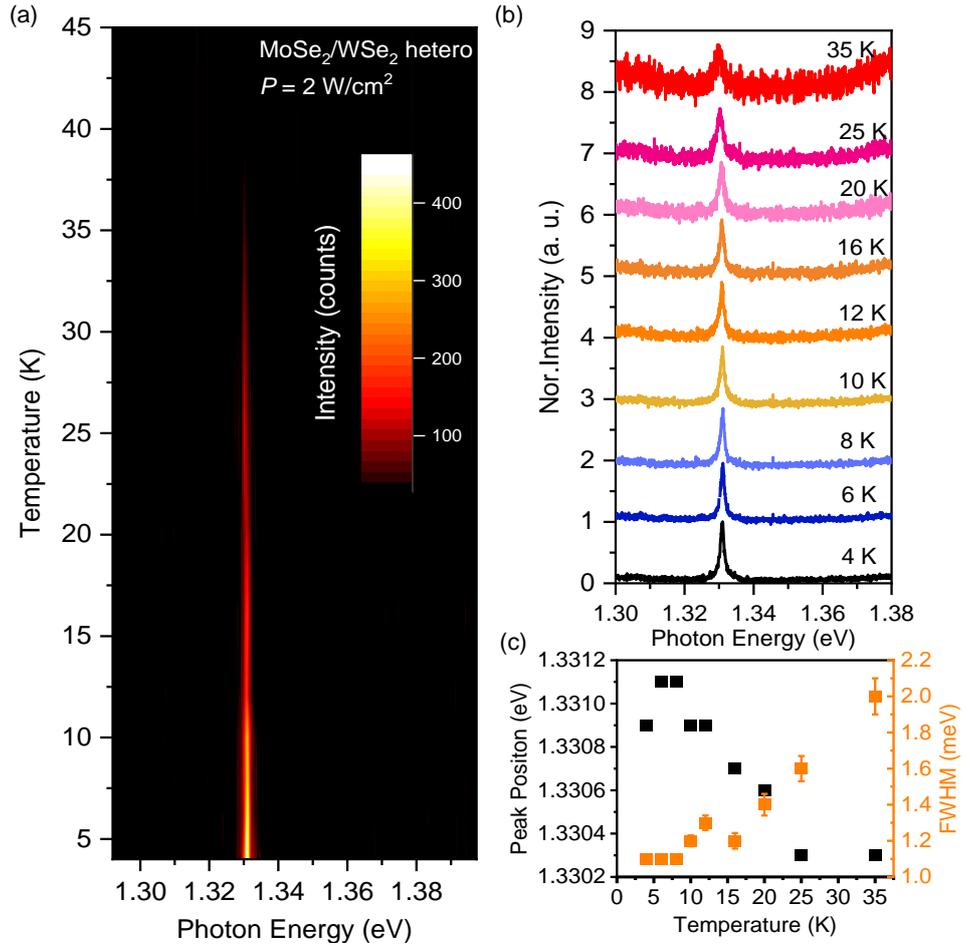

**Supplementary Figure 4: Temperature dependence of PL spectra of a single moiré exciton measured under power density of 0.56 W/cm². (a)** Contour plot of the PL spectrum from 4 to 45 K. **(b)** Normalized PL spectra measured under various temperatures. **(c)** Peak position and spectral linewidth defined by full-width at half-maximum (FWHM) as a function of temperature.

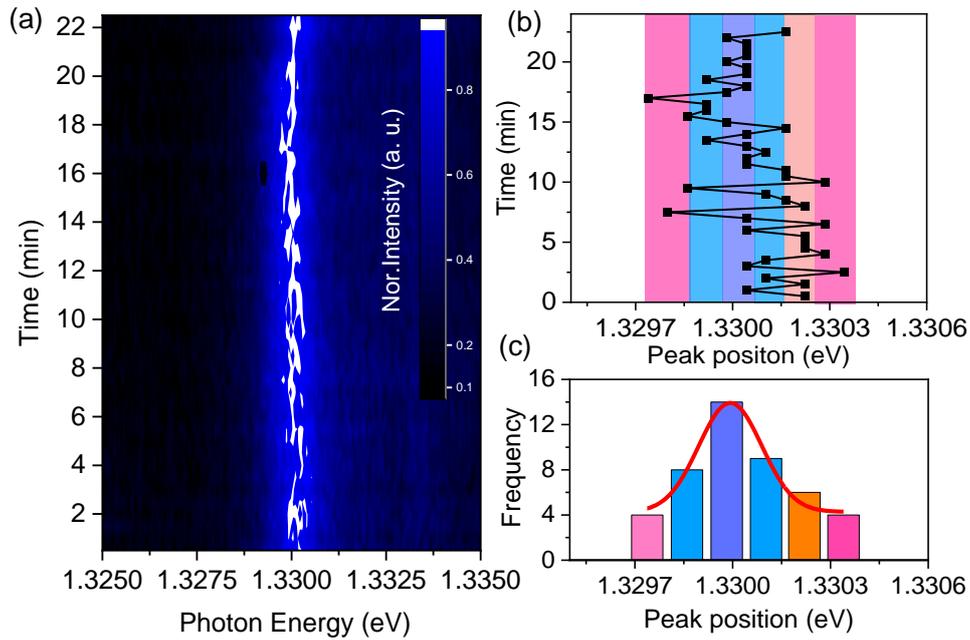

**Supplementary Figure 5: Time-trace of PL spectra of a single moiré exciton measured under 4 K and excitation power density of 34 W/cm$^2$.** **(a)** PL spectra of a single moiré exciton in the heterobilayer. Each spectrum is measured during an accumulation time of 30 seconds. The time-trace PL spectra show spectral wondering. **(b)** Time-trace of spectral peak positions recorded from the spectra. The color bars represent the energy windows with a constant width, where the frequencies of peak positions are counted. The distribution range of peak position is 0.6 meV from 1.3297 to 1.3303 eV. **(c)** Frequency of the peak positions counted in each energy windows. The distribution of frequencies in the energy peak positions can be fitted with Gaussian function, indicated by the red solid curve. The center and width of Gaussian distribution are 0.2 meV and 1.3300 eV, respectively.

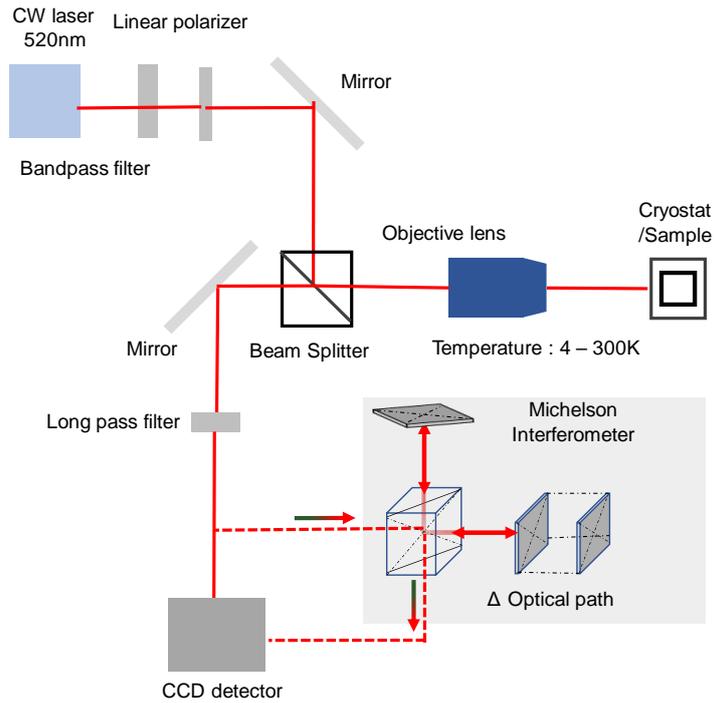

**Supplementary Figure 6: Optical setup for the low temperature PL measurement and coherence measurement.** In the measurement of the first-order correlation function, Michelson interferometer is inserted before the detector.

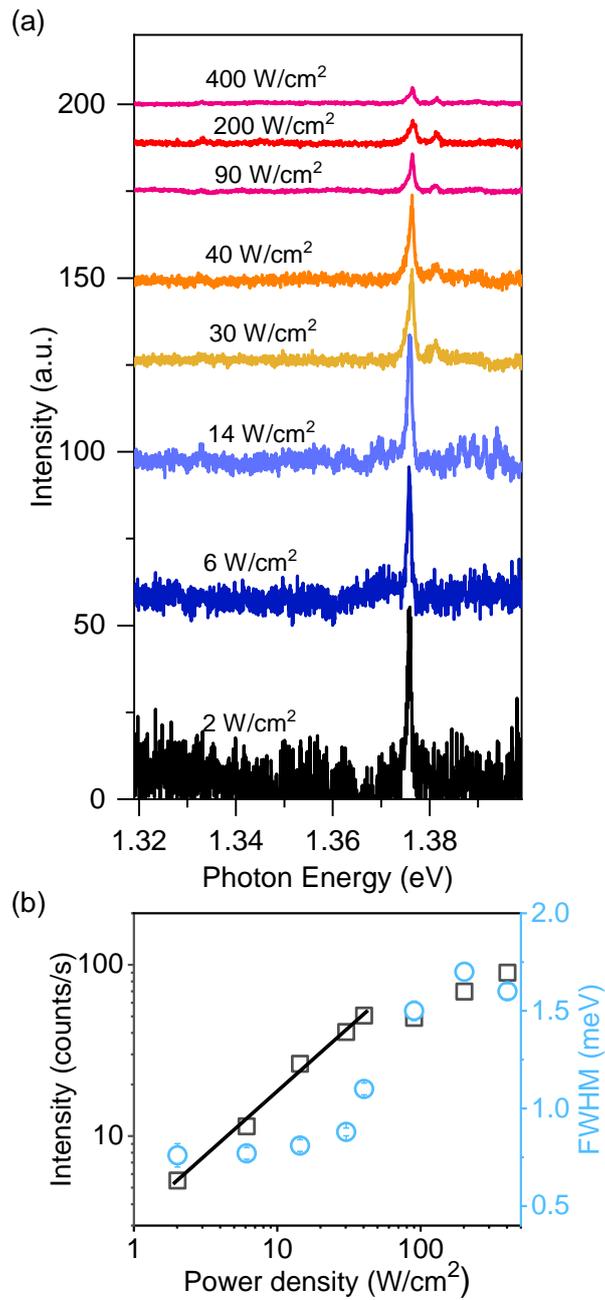

**Supplementary Figure 7: PL spectra measured at 4 K under various power densities (a)** Power dependence of PL spectra measured under 4 K in the heterobilayer. **(b)** Integrated PL intensity and linewidth defined by FWHM, indicated by black sloid squares and sky-blue circles as a function of the excitation power density. The solid line shows the guide for linear power dependence.

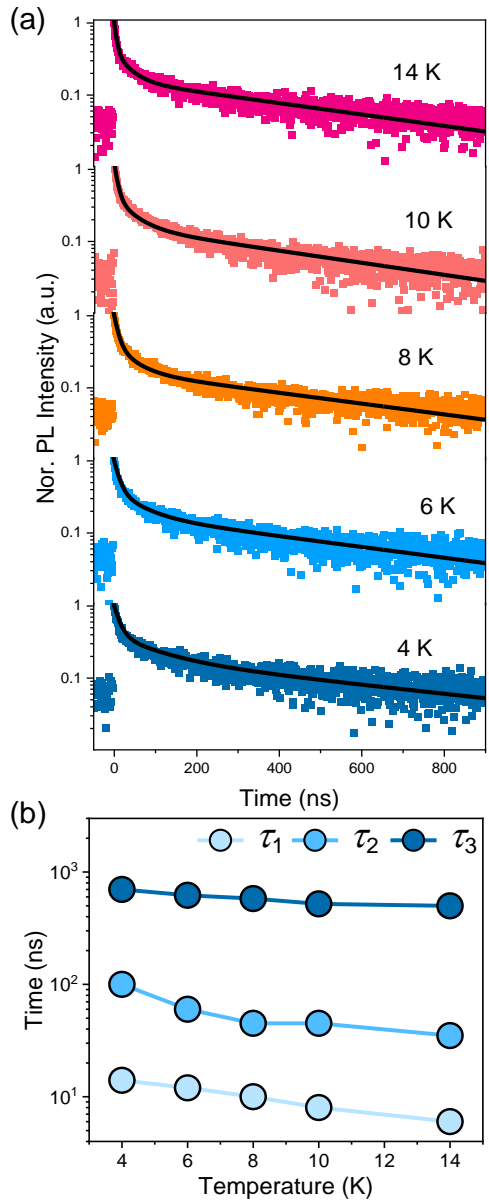

**Supplementary Figure 8: Temperature dependence of PL decay profiles of moiré exciton under excitation power density of 28 W/cm$^2$ (a)** Time evolutions of PL intensity of moiré exciton measured from 4 to 14 K. A supercontinuum laser with repetition rate of 1 MHz filtered by 1.72 eV band pass filter was used. The multi-exponential functions are used for the fitting of data, presented by the black solid curve. **(b)** PL lifetimes obtained from the fitted results as a function of temperature.

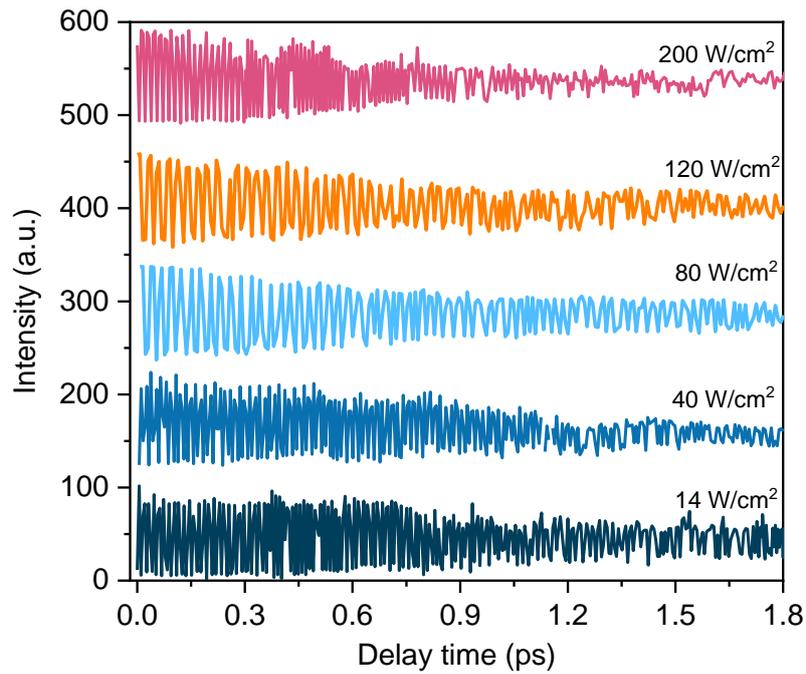

**Supplementary Figure 9: Interferogram of a single moiré exciton measured at 4 K under various excitation power densities.** The fringe represents the maximum and minimum PL intensity at various delay time. Decays of fringe visibility obtained from the maximum and minimum intensity as a function of the delay time are plotted in Fig. 3g.

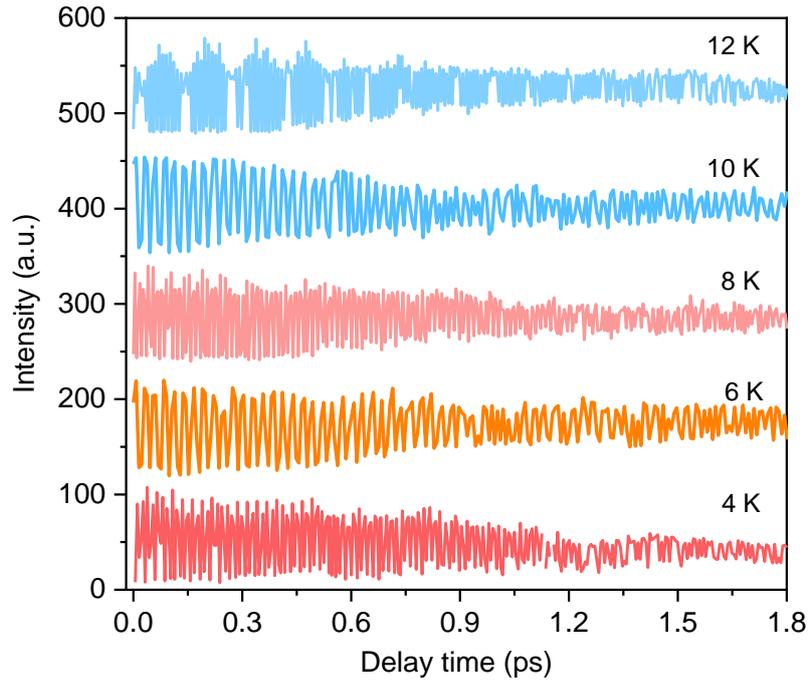

**Supplementary Figure 10: Interferogram of a single moiré exciton measured under excitation power density of 40 W/cm$^2$ at various temperatures.** Decays of fringe visibility obtained from the maximum and minimum intensity as a function of the delay time are plotted in Fig. 3i.

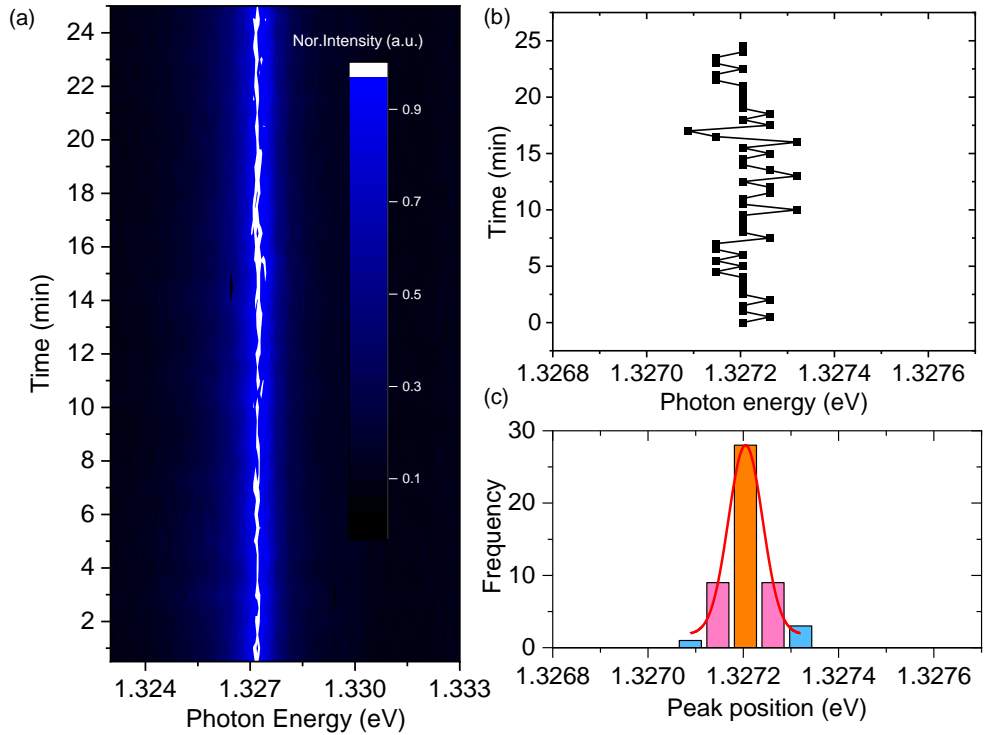

**Supplementary Figure 11: Time-trace of PL spectra of a single moiré exciton measured under 4 K and excitation power density of 3.2 W/cm$^2$.** **(a)** Time trace of PL spectra of a single moiré exciton. Each spectrum is measured during an accumulation time of 30 seconds. The time-trace PL spectra show smaller spectral wondering in Fig. S5. **(b)** Time-trace of spectral peak positions recorded from the spectra. The result shows that the distribution range of peak position is 0.2 meV from 1.3271 to 1.3273 eV, which is smaller than the range from previous results. **(c)** Frequency of the peak positions counted at each energy window. The distribution of frequencies in the energy peak positions can be fitted with Gaussian function, indicated by the red solid curve. The center and width of Gaussian distribution are 0.07 meV and 1.3272 eV, respectively.